\documentclass{sig-alternate}
\usepackage{url}
\usepackage{algorithm}
\usepackage{algorithmic}

\newtheorem{dfn}{Definition}
\begin{document}

\title{Predicting Popularity of
Twitter Accounts through the Discovery of Link-Propagating Early
Adopters}

\numberofauthors{2}%
\author{
\alignauthor
Daichi Imamori%
\\
       \affaddr{Graduate School of Informatics, Kyoto University}\\
       \affaddr{Sakyo, Kyoto 606-8501 Japan}\\
       \email{imamori@dl.kuis.kyoto-u.ac.jp}
\alignauthor
Keishi Tajima\\
       \affaddr{Graduate School of Informatics, Kyoto University}\\
       \affaddr{Sakyo, Kyoto 606-8501 Japan}\\
       \email{tajima@i.kyoto-u.ac.jp}
}

\maketitle
\begin{abstract}
In this paper, we propose a method of ranking recently created Twitter
accounts according to their prospective popularity.  Early detection
of new promising accounts
is useful for trend prediction, viral marketing, user recommendation,
and so on.  New accounts are, however, difficult to evaluate because
they have not established their reputations, and we cannot apply
existing link-based or other popularity-based account evaluation
methods.  Our method first finds ``early adopters,'' i.e., users who
often find new good information sources earlier than others.  Our
method then regards new accounts followed by good early adopters as
promising, even if they do not have many followers now.  In order to
find good early adopters, we estimate the frequency of link
propagation from each account, i.e., how many times the follow links
from the account have been copied by its followers.  If its followers
have copied many of its follow links in the past, the account must be
an early adopter, who find good information sources earlier than its
followers.  We develop a method of inferring which links are created
by copying which links.  One advantage of our method is that our
method only uses information that can be easily obtained only by
crawling neighbors of the target accounts in the current Twitter
graph.
We evaluated our method by an experiment on Twitter data.  We chose
then-new accounts from an old snapshot of Twitter, compute their
ranking by our method, and compare it with the number of followers
the accounts currently have.  The result shows that our
method produces better rankings than various baseline methods,
especially for new accounts that have only a few followers.
\end{abstract}



\section{Introduction}
\label{sec:intro}

In social media, such as micro-blogs and social network services,
users can easily create new accounts and quickly start up new
information publishing channels at low cost.
As a result, social media are highly dynamic world.  Micro-blogging
services, such as Twitter, are especially dynamic because they focus
more on prompt
information dissemination, while social network services, such as
Facebook, focus more on communication over longer-term social
relationship.

Because of the dynamicity, new popular accounts continually appear and
disappear in micro-blogging services.  Early detection of new accounts
that will become popular in future is an important problem that has
several applications, such as trend detection, viral marketing, and
user recommendation.

Estimation of popularity of an account is also useful for
approximating the quality of information it posts.  Estimation of the
quality of information is very important in many applications, but it
is generally difficult to estimate it without human intervention.  To
solve this problem, popularity-based methods have been widely used.
Methods that estimate information quality of web pages based on the
number of their incoming links
has been successful \cite{KleinbergHITS,BrynPageRank}.  Similar idea
has also been successfully applied to micro-blogs with linking
functions \cite{weng2010twitterrank}.  These facts proved that there
is high correlation between the popularity and the quality of
information.  Therefore, the estimation of prospective popularity of
new accounts, which have not yet established the popularity they
deserve, is also useful for estimation of the quality of new
information sources.

In this paper, we propose a method of predicting future popularity of
new Twitter accounts, in other words, the number of followers they
will obtain in future.

\subsection{Our Approach}

The most important factor deciding the future popularity of an account
is, of course, the quality of information it posts, but it is even
more difficult to estimate as explained above.  That is exactly one of
the reasons why we want to predict popularity instead.  We therefore
should explore a method of predicting future popularity of an account
not based on its information quality but based on its current
popularity.

New accounts, however, usually have only a small number of followers.
How to predict future popularity only with that information is the
challenge of the problem we discuss in this paper.  Because the number
of followers is usually small, we also use the quality of each
follower.  It is basically the same approach as many existing
link-based quality estimation methods
\cite{KleinbergHITS,BrynPageRank,weng2010twitterrank}.

We focus on a specific type of quality of followers that is most
important for our purpose: whether the link from it implies more links
in future.  In Twitter, and in social media in general, there are
users that are good at finding good information sources earlier than
other users.  We call such users \emph{early adopters}.  Early
adopters themselves often have many followers, and when an early
adopter creates a link to a new good information source, many of its
followers imitate it and create links to the information source.  In
other words, early adopters play the role of hubs for link propagation
in social media, and therefore, links from good early adopters imply
more links in future.

Our method predicts future popularity of new accounts by estimating
how good their current followers are as early adopters.  If a new
account is followed by good early adopters, our method regards the new
account as promising, even if it does not have many followers now.

Our method determines whether a given user is a good early adopter by
estimating the frequency of link propagation through it in the past,
i.e., how many times its links have been imitated by its followers.
If links from it has been imitated by its followers many times, the
user must be a good early adopter who can find good information
sources earlier than its followers.

\subsection{How to Detect Copied Links}

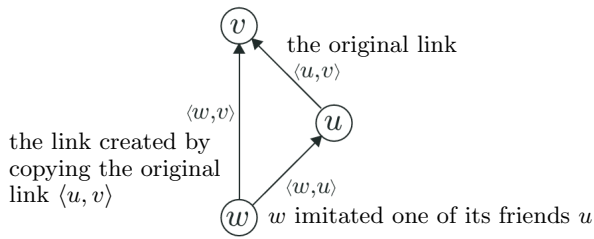
\begin{figure}[t]
\centering
\includegraphics[scale=0.25,bb=0 0 250 347]{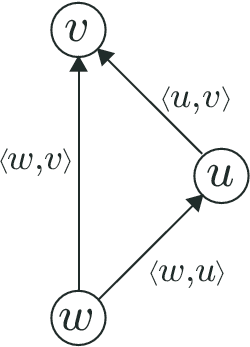}
\put(-25,70){the original link}
\put(-130,33){the link created by}
\put(-130,23){copying the original}
\put(-130,13){link $\langle u,v\rangle$}
\put(-32,5){$w$ imitated one of its friends $u$}
\caption{Triadic closure produced by user $w$'s copying of $u$'s
follow link to $v$.}
\label{fig:triad}
\end{figure}

In Twitter, however, the information on which link was created by
copying which link is not immediately available.  We infer it by using
several factors.
The most important factor is network structure.
We assume that link propagation occurs only from a user to its
followers.  In other words, each user only imitates links of his/her
friends (users that he/she follows).  If we assume this, a link
created by imitation must be a part of a triangle consisting of three
links: an original link, a link created by copying it, and a link from
the user who copied the link to the user whose link was copied.
Figure~\ref{fig:triad} shows an example of such a triangle.  We first
collect candidates of links created by imitation by finding triangles
that may correspond to such structure.  We call such candidate
triangles \emph{triadic closures}.

In Twitter, users very often find new information sources by browsing
the friend lists of their friends and copying some of them which seem
interesting to them.  This kind of practice is not specific to
Twitter, and is quite common to many social media.  It is one of key
differences between social media and other older media, such as RSS
(RDF Site Summary or Really Simple Syndication) \cite{rss2.0}, where
users cannot browse other users' subscription.
We also think it is one of the feature that promoted the growth of
social media like Twitter over the older media.  In addition, in
Twitter, users can ``retweet'' (i.e., forward) a tweet from their
friends to their followers, and when users find a tweet retweeted by a
friend interesting to them, they often create direct follow links to
the account that originally posted the tweet.  Similar forwarding
functions are found in many social media.

These observations are the rationale of our assumption that link
propagation mainly occurs from users to their followers.  Inclusion of
other kinds of link propagation or link creation into the model is an
interesting direction for future research.

In order to further narrowing down the candidates of links created by
imitation, we also consider three other factors: time order of link
creation, link reciprocity, and the similarity between users.

The first factor is time order of link creation.  In a triadic
closure, the link created by copying must be newer than the other two
links.  Otherwise, it must not be a result of copying.

The second factor, reciprocity of links, is used for distinguishing
links to information sources and links between personal friends.
In Twitter, non-reciprocal links are more likely to refer to
information sources than reciprocal links are
\cite{XieFriendTwitterLife}.  Because links to information sources are
more important for the discovery of early adopters, we distinguish the
two types of links based on their reciprocity.

\begin{figure}[t]
\centering
\includegraphics[scale=0.25,bb=0 0 320 347]{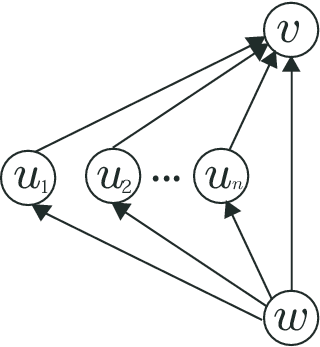}
\put(0,50){a link created}
\put(0,40){by copying}
\put(-130,70){multiple candidates}
\put(-130,60){of the original}
\caption{The follow link from $w$ to $v$ is a part of many triadic
closures and it is not obvious who in $u_1,\ldots,u_n$ was imitated by
$w$.}
\label{fig:rival}
\end{figure}
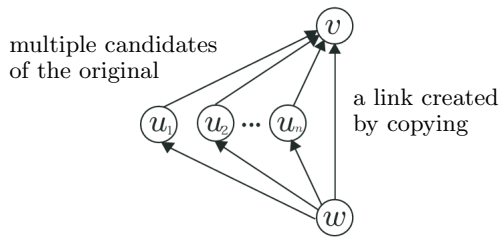

Even if we find a triadic closure and the links in it satisfy the
conditions above, the candidate link in it may not actually be a copy
of the link in that triad.  If the candidate link is also a component
of many other triadic closures, it may be a copy of another link in
another triadic closure.  Figure~\ref{fig:rival} illustrates such a
situation.  In this example, the follow link from the user $w$ to the
user $v$ is a part of many triadic closures, and it is not obvious who
in $u_1,\ldots,u_n$ was imitated by $w$.

We estimate the probability that a candidate link is a copy of the
link in a given triad by using similarity between interests of users
in the triad.  This is the third factor.  This factor is based on an
assumption that link propagation is more likely to happen when the
interests of related users are similar to each other.  We measure the
similarity of interests of users by the similarity of their friend
lists.

We use these three factors as optional factors, and tested all eight
combinations of them in our experiments.  Our experimental result
shows that the link reciprocity is very useful for improving the
accuracy of our method, but the other two are not very useful.  The
details will be explained later.

Notice that all information used by our method, network structure and
the three optional factors explained above, can be obtained easily
only by crawling the neighbors of the target new accounts in the
current Twitter graph structure.  This is one important advantage of
our method.

In Twitter, link propagation to followers is especially likely to
occur when users receive interesting messages
retweeted
by their friends.  Our method, however, does not use the information
on retweeting because it requires monitoring of the tweet stream, and
we would lose the advantage of our method mentioned above.

By using network structure and the three optional factors above, we
infer which links are copies of which links, and we estimate how many
times each user has been imitated by other users.  Based on this
value, we compute \emph{early adopter score} of each user.  We then
compute \emph{future popularity score} of each account based on the
early adopter scores of its followers.

\subsection{Comparison with Baseline Methods}

We conducted an experiment with a data set consisting of a part of
Twitter graph that were collected by a random crawler on May 2012
\cite{DBLP:conf/kdd/LiWDWC12}.  We estimated future popularity of
then-new accounts in this data set, and compared the result with the
number of non-reciprocal followers that the accounts currently have as
of May 2015, which indicates how popular these accounts are now.  The
result of the experiment shows that our method outperforms various
baseline methods when we compare the accuracy of the whole ranking of
all the new accounts.  Our method outperforms baseline methods
especially when we apply them to new accounts that have only a few
followers.

In addition, the correlation between the ranking by our method and
those by baseline methods are low.  This suggests that our method and
baseline methods are complementary to each other.  Our experimental
results shows that we can actually produce a even better ranking by
logistic regression combining our method and some of the baseline
methods.

When we compare the accuracy of only the top part of the rankings, a
variation of HITS algorithm \cite{KleinbergHITS} or a variation of
PageRank method \cite{BrynPageRank}
achieves the highest accuracy in most cases.
Even a naive method that ranks the accounts by the number of the
non-reciprocal followers they had at that time ourperforms our method
in some cases.  It is mainly because the top part of the rankings
include accounts that were already popular in the old snapshot.  This
fact again suggests that our method is particularly useful when we
want to find new accounts that are not popular now but will be popular
in future.

Although our main purpose is to predict future popularity of new
accounts, we also expect that analysis of early adopters discovered by
our method would also help revealing what factors are important for
new accounts to obtain links from early adopters, and what factors are
important for users to be good early adopters.

\section{Related Work}
\label{sec:related}

In sociology, there has been extensive research on the behavior of
people in the real world.  The results of the research are also
helpful for understanding the behavior of people on social media.
There have been some studies that have shown that the behavior
reported in the past research in sociology is also observed on social
media
\cite{hu2012people,nguyen2010you,hopcroft2011will,romero2010directed}.
One of such studies on the behavior of people has proposed the concept
of triadic closure for explaining how people behave when they connect
to each other.
Our method uses this concept for inferring which links are created by
copying which links.

Recently, there have also been many studies on link prediction in
online social network.  For example, Liben-Nowell and Kleinberg
\cite{liben2007link} was the first to formulate the problem of
link-prediction on social network and they proposed a prediction
method based on the proximity of nodes in the network.  Zhang et al.\
\cite{zhang2013learning} proposed a method that estimates the
probability of future links by inferring latent paths of link
propagation in the network.  They estimate how important each node is
as a mediator of link propagation by using a probabilistic model.  Our
method is based on a similar concept of early adopters.  Their method,
however, requires multiple snapshots of the network structure at
different time point.  On the other hand, our method for estimating
the future popularity of a given account only requires information
that can be obtained by crawling the neighbors of the target account
in the current snapshot of the network structure.  This is one big
advantage of our method.

There have also been many studies on estimation of the influential
power of nodes in social network.  For example, Kwak et al.\
\cite{kwak2010twitter} compared three indicators, PageRank, the number
of followers, and the number of retweets, for the estimation of
popularity of Twitter accounts, and they showed that there is a
discrepancy between the number of followers of an account and the
popularity of tweets by the account, which suggests that the number of
followers is not an only major factor of influential power of nodes.
Weng et al.\ \cite{weng2010twitterrank} also proposed a method for
estimating influential power of Twitter accounts.  Their method is
based on the number of followers, but they also consider the interests
of the followers and compute the probability that each tweet is
actually read by the followers.  These two studies focus on
influential power of information sources in information dissemination,
while the early adopter score used in our method indicates the
influential power of nodes in link propagation.

The discovery of early adopters in online community has been discussed
in several studies.  Bakshy et al.~\cite{BakshySocialInfluence}
analyzed how users adopt new contents in a social network in Second
Life, and identified early adopters, but also found that early
adopters do not always have significant influence on the other users.
Saez-Trumper et al.~\cite{SaezTrumperTrendsetters} proposed a method
of identifying early adopters that also have significant influence on
the others in information network, such as Twitter, and called such
users trendsetters.  These studies focused on temporal relationship of
users' adoption of contents, such as hashtags and URLs.  Goyal et
al.~\cite{GoyalDiscoveringLeaders} also proposed a method of
identifying leaders in online communities whose actions, e.g., tagging
resources or rating songs, are imitated by many users.  On the other
hand, we focused on the adoption of new Twitter accounts, i.e., the
creation of new follow links, and imitation of them by the followers.
We showed that the idea similar to theirs can also be applied to such
a type of actions in order to predict future popularity of new
accounts in Twitter.  Another contribution of this paper is to develop
a method of inferring who copied which follow links, the information
which is not immediately available.

\section{Scores for Prediction}
\label{sec:propagation}

In this section, we
define future popularity scores of accounts, which we use for ranking
new accounts based on its prospective popularity, and also define
early adopter scores of accounts, which we use for computing future
popularity scores of their friends.
We first define notation used in this paper.  Let $G(V, E)$ be the
follow graph of Twitter, where $V$ is a set of all Twitter accounts,
and $E$ is a set of all follow links among them.  $\langle u,
v\rangle\in E$ denotes a follow link from a user $u$ to a user $v$.
For $u \in V$, $\mathit{Friends}(u)$ denotes the set of users followed
by $u$, and $\mathit{Followers}(u)$ denotes the set of followers of
$u$.  Similarly, $\mathit{Friends}(S)=\bigcup_{u\in S}\mathit{Friends}(u)$ and
$\mathit{Followers}(S)=\bigcup_{u\in S}\mathit{Followers}(u)$.

\subsection{Early Adopter Score: {\large\boldmath $E$}}
\label{sub:earlyadopter}

We first define early adopter scores of accounts.  Let
$\mathit{Copy}(u)$ denote a set of links $\langle w, v\rangle$ which a
user $w$ who is a follower of $u$ created by copying a link $\langle
u, v\rangle$.  Figure~\ref{fig:triad} illustrates an example of such a
link structure.  $|\mathit{Copy}(u)|$ takes its maximum value when all
followers of $u$ imitated all links by $u$.  Therefore,
$|\mathit{Copy}(u)| \leq |\mathit{Followers}(u)| \times
|\mathit{Friends}(u)|$.
We then define $I(u)$, the imitation ratio of $u \in V$, as follows.
\begin{dfn}
The imitation ratio of $u$:
\[
I(u) = \frac{|\mathit{Copy}(u)|}{|\mathit{Followers}(u)| \times |\mathit{Friends}(u)|}
\]
\end{dfn}
The numerator is the number of times $u$ was imitated by its
followers.  The denominator is the maximum value that the numerator
can take.  When the denominator is $0$, we let $I(u)=0$.  $I(u)$
represents the probability that a link of $u$ is imitated by its
followers.  How to infer $\mathit{Copy}(u)$ based on network structure
and three optional factors will be explained later.

Based on $I(u)$, we define the early adopter score of $u$.
We define it in two ways, and compare their performance by the
experiment later.  Both definitions try to estimate the expected
number of link propagation through $u$, but they are based on
different assumptions.

The first definition is based on the following assumption.  Suppose a
new information source $v$ is newly followed by an early adopter $u$.
We then expect that each of the follower of $u$ will follow $v$
independently in the probability $I(u)$.  Therefore, the expected
number of new follow links created by imitating $u$ is
$|\mathit{Follower}(u)|\times I(u)$.

However, we predict future popularity of an account based on the
current snapshot.  Even if a recently created account $v$ is followed
by an account $u$ in the snapshot, if most followers of $u$ already
have links to $v$ in the snapshot, we cannot expect that many users
will newly follow $v$ by imitating $\langle u,v\rangle$.  In other
words, $u$ is not an ``early adopter'' compared with its followers
with respect to $v$.  With including this factor in the computation,
we define $E_1(u,v)$, the first variation of an early adopter score of
$u$ with respect to $v$, as follows.

\begin{dfn}
The early adopter score of $u$ with respect to $v$ (variation 1):
\[
E_1(u,v) = I(u)\times |\mathit{Followers}(u) \setminus \mathit{Followers}(v)|
\]
\end{dfn}
This represents the expected increase of the number of followers of
$v$ through $u$.

The second definition of the early adopter score of $u$ is based on
the following assumption.  Suppose an information source $v$ is
followed by an early adopter $u$ in the current snapshot.  Some of the
followers of $u$ already have links to $v$.  The other followers of
$u$ are not likely to follow $v$ from now because they have not done
so until now.  However, the new followers that $u$ will obtain from
now will follow $v$ by imitating $u$ in the probability $I(u)$.  The
number of followers that $u$ will obtain from now are unknown, and we
simply assume that it is a constant $n$ for any $u$.  Under this
assumption, the expected increase of the number of followers of $v$
through $u$ is $n\times I(u)$.  Because we use early adopter scores
for computing ranking scores of accounts, we can ignore the constant
$n$, and we define the second variation of the early adopter score as
follows.
\begin{dfn}
The early adopter score of $u$ (variation 2):
\[
E_2(u,v) = I(u)
\]
\end{dfn}
The second parameter $v$ of $E_2(u,v)$ is used only for the
compatibility with the first variation $E_1(u,v)$, and is not actually
used in this second variation.

There is another way to interpret $E_2(u,v)$.  $I(u)$ represents how
good $u$ is as an early adopter.  If a new account $v$ is followed by
a good early adopter, we can expect that the quality of $v$ is high,
and therefore, we can expect that it will have many followers in
future, no matter these new followers would find $u$ through $v$ or
not.  Therefore, we can simply use $I(u)$
for computing future popularty scores of accounts followed by $u$.

\subsection{Future Popularity Score: {\large\boldmath $F$}}
\label{subsec:eus}

By using the early adopter score defined above, we next define the
future popularity score of an account $v$.  The simplest way to define
it is to sum up the early adopter scores of all the followers of $v$:
\begin{dfn}
Sum-based future popularity score of $v$:
\[
F_i^{\Sigma}(v) = \sum_{u \in \mathit{Followers}(v)}E_i(u, v)
\]
\end{dfn}
where $i$ is either 1 or 2.  This definition, however, has a problem
when we use $E_1$.  $E_1(u,v)$ represents the expected increase of the
followers of $v$ through $u$, and if $u_1$ and $u_2$ have some common
followers, simply summing up $E_1(u_1,v)$ and $E_1(u_2,v)$ would
double-counts those common followers.  Therefore, we should define
$F_1^{\Sigma}(v)$ in the following way:
\begin{dfn}
Sum-based future popularity score of $v$ (the second definition):
\[
F_1^{\Sigma}(v) = \sum_{w\in \mathit{Followers}(\mathit{Followers}(v))}P(\bigvee_{u\in \mathit{Followers}(v)}cp(w, \langle u,v\rangle))
\]
\end{dfn}
where $P(e)$ is the probability of the event $e$ and $cp(w, \langle
u,v\rangle)$ is the event that $w$ copies the link $\langle
u,v\rangle$.  That is, we sum up the probability that a follower $w$
of some follower of $v$ will follow $v$ by imitating any of the
followers of $v$.  We compute this probability by assuming that events
$cp(w, \langle u_i,v\rangle)$ and $cp(w, \langle u_j,v\rangle)$ are
independent for $i\neq j$ and $P(cp(w, \langle u,v\rangle)) = I(u)$.
According to our experiment, however, the performance of
$F_1^{\Sigma}(v)$ in this definition and that of the previous simpler
definition have no significant difference.  Therefore, we use the
previous simpler definition for both $F_1^{\Sigma}(v)$ and
$F_2^{\Sigma}(v)$.

A disadvantage of these sum-based definitions of future popularity
scores is that it basically gives higher scores to accounts with many
followers.  Our purpose is to evaluate future popularity of new
accounts that have not obtained many followers.  Giving higher scores
to accounts that already have many followers contradicts to our
purpose.

Another way to define future popularity scores with avoiding that
problem is to use g-index \cite{egghe2006theory} instead of sum in the
following way.

\begin{dfn}
G-index-based future popularity score of $v$:
\[
F_i^{g}(v) = \mbox{RG}(\{E_i(u, v)\ |\ u\in \mathit{Followers}(v)\})
\]
\end{dfn}
where $i$ is either $1$ or $2$ and $\mbox{RG}(S)$ is a function that
computes the rational g-index of the set of real numbers $S$
\cite{tol2008rational}.

$F_i^{g}(v)$ is a rational g-index of the set of the early adopter
scores of the followers of $v$.  Given a set $S$ of values, its
g-index can be computed by the following procedure.  First we make a
list $L$ by sorting values in $S$ in decreasing order.  Let $L[i]$ be
the $i$-th value in $L$.  We then find a maximum $g$ that satisfies
$g^2\leq c\times\sum_{i \leq g}L[i]$, where $c$ is a parameter, and
such a $g$ is the g-index of $S$.
G-index of a set $S$ is affected only by largest values in $S$.
G-index of a set only takes natural numbers, but rational g-index is
an extension of g-index to rational numbers \cite{tol2008rational}.

\section{Detection of Link Imitations}
\label{sec:method}

In the previous section, we defined early adopter score and future
popularity score.  In order to compute these scores, however, we first
need to know $\mathit{Copy}(u)$, i.e., the set of links created by imitating a
link from $u$.  This information is not immediately available from
Twitter data.  In this section, we propose a method to estimate
$|\mathit{Copy}(u)|$, i.e., the number of times such imitation has occurred.
As explained in Section~\ref{sec:intro}, we use network structure and
the following three optional factors for it:
\begin{itemize}
\item temporal order of creation of links
\item link reciprocity
\item similarity among interests of users
\end{itemize}

\subsection{Network Structure}

Network structure is the most basic information for collecting
candidates of links created by imitation.  In this paper, we assume a
user copies follow links only from his/her friends.  On this
assumption, if a link $\langle w, v\rangle$ is a copy of $\langle u,
v\rangle$, there must also be a link $\langle w, u\rangle$, i.e., they
must form a triangle shown in Figure~\ref{fig:triad}.  We call such a
triangle
\emph{triadic closure} \cite{rapoport1953spread}.  We first find such
triadic closures.

We define a boolean function $\mathit{Structure}(u, v, w)$ that
determines if $u, v, w \in V$ form a triangle that could be a triadic
closure as follows:

\begin{dfn}
Function determining if $u, v, w \in V$ of $G(V,E)$ forms a valid
triangle:
\[
\mathit{Structure}(u, v, w) =
(\langle u, v\rangle\in E)\land (\langle v, w\rangle\in E)\land (\langle w, u\rangle\in E)
\]
\end{dfn}

\subsection{Time Order of Link Creation}

Triangles satisfying the condition above do not necessarily correspond
to triadic closures.  We can further narrow down the candidates of
triadic closures by using three optional factors.  The first factor is
temporal order of creation of links.  In the triangle satisfying the
condition above, $\langle w,v\rangle$ must be newer than $\langle
u,v\rangle$ and $\langle w,u\rangle$ if $w$ created it by copying
$\langle u,v\rangle$.

This information can be retrieved from the current Twitter data.
Twitter API provides functions that return a list of followers and a
list of friends of a given user.  These functions return lists sorted
by time when they became followers or friends from the newest one to
the oldest one.  Let $idx(v,l)$ denotes the position of $v$ in the
list $l$.  A boolean function representing whether the triangle
satisfies the necessary temporal condition is then defined as follows:

\begin{dfn}
Condition on temporal order of creation of links in a candidate
triangle consisting of $u, v, w\in V$:
\begin{eqnarray*}
\lefteqn{\mathit{Time}(u, v, w) =}\\
&& idx(v,\mathit{Friends}(w)) < idx(u,\mathit{Friends}(w))\ \land\\
&& idx(w,\mathit{Followers}(v)) < idx(u,\mathit{Followers}(v))
\end{eqnarray*}
\end{dfn}

\begin{figure}[t]
  \centering
  \includegraphics[scale=0.25,bb=0 0 855 390]{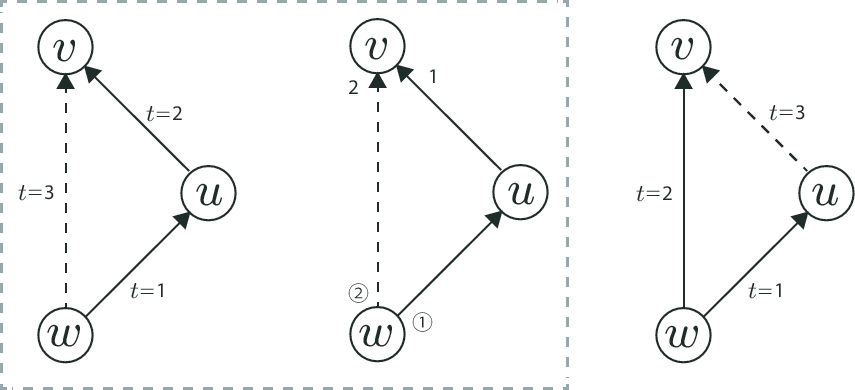}
  \caption{In the triad at left, $\langle w,v\rangle$ may be a copy of
  $\langle u,v\rangle$, but in the triad at right, $\langle
  w,v\rangle$ never be.  The condition can be determined by the order
  of $u,v$ in the friend list of $w$ and the order of $u,w$ in the
  follower list of $v$, as shown at middle.}
  \label{fig:temporalOrder}
\end{figure}

Figure~\ref{fig:temporalOrder} illustrates examples of valid triangle
(left) and invalid triangle (right).  It also shows how we can check
the conditions by the positions of $u$, $v$, $w$ in $\mathit{Friends}(w)$ and
$\mathit{Followers}(v)$.

One disadvantage of this optional condition is that we need to store
the time order of friends of $w$ and followers of $v$.  In addition,
surprisingly, this condition does not improve the performance of our
method much, as explained later in Section~\ref{sec:experiment}.

\subsection{Reciprocity of Links}
\label{subsec:nonreciprocalFollow}

Follow links in Twitter can be classified into several types, such as
links to information sources and links to personal friends.  There is
also a practice called \emph{followback}.  In Twitter, some users
follow back to many of its followers as an act of courtesy.
Among these three types of links, the latter two are usually
reciprocal.  Personal friends usually link to each other
\cite{XieFriendTwitterLife}, and links created by followback are
always reciprocal.  On the other hand, links to information sources
are usually non-reciprocal unless the information source is a type of
users who always follow back to all its followers.

In the following, $\mathit{Followers_{nr}}(u)$ denotes the set of
non-reciprocal followers of $u$, i.e.,
$\mathit{Followers_{nr}}(u)=\mathit{Followers}(u)\setminus\mathit{Friends}(u)$.
Similarly, $\mathit{Friends_{nr}}(u)$ denotes the set of
non-reciprocal friends,
$\mathit{Friends_{nr}}(u)=\mathit{Friends}(u)\setminus\mathit{Followers}(u)$.

For the discovery of early adopters, links to information sources are
important.  Therefore, we should exclude the other types of links from
the consideration in our method.  Although it is difficult to fully
distinguish links to information sources from the others,
we may be able to improve the precision of our method by excluding (or
by giving lower weights to) reciprocal links because it excludes most
of the other types of links (while it also excludes some links to
information sources).  Our experimental result, which will be shown in
Section~\ref{sec:experiment}, shows that we can actually improve the
precision by excluding reciprocal links.

\begin{figure}[t]
  \centering
  \begin{picture}(120,110)(0,0)
  \put(0,0){\includegraphics[scale=0.3,bb=0 0 366 364]{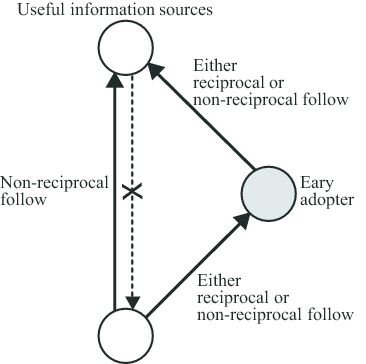}}
  \put(77,48){{\Large $u$}}
  \put(34,92){{\Large $v$}}
  \put(33,5){{\Large $w$}}
  \end{picture}
  \caption{Exclusion of reciprocal links from candidates.}
 \label{fig:reciprocity}
\end{figure}
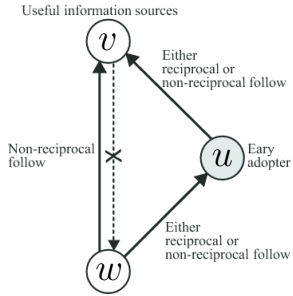

Based on the discussion above, we define $\mathit{Nonrec}(u, v, w)$,
the weight of the triad consisting of $u,v,w$, by the formula below:
\begin{dfn}
The weight given to a triad of $u, v, w$:
\[
\mathit{Nonrec}(u, v, w) = v \in \mathit{Friends_{nr}}(u)
\]
\end{dfn}
In this paper, we simply assign the weight 0 to triads that have
reciprocal links between $w$ and $v$.  In other words, we exclude
reciprocal links from the candidates of links created by imitation.
Figure~\ref{fig:reciprocity} illustrates this condition.
By this factor, we expect that we can distinguish triadic closures
corresponding to a circle of personal friends and those corresponding
to imitation of links of early adopters.

\subsection{Similarity between Interests of Users}

By using the three conditions above, we can narrow down candidate
triads corresponding to link imitation.  However, if there are
multiple candidates of the original link for one link, only one of
them was really copied.  When we have such multiple candidates,
instead of selecting one of them as the original link, we assign each
of them the probability that it is really the imitated one.

The simplest way to assign the probability is to assign equal
probability to all the candidates.  We also designed another way to
assign probability that is proportional to the similarity between
interests of users, which is the third optional factor.

In Twitter, various users with various interests publish or collect
information.  Early adopters must also have some specific interests,
and each early adopter must be good at finding new useful information
only on some topics.  Similarly, users imitating early adopters also
have some specific interests, and they are more likely to imitate
early adopters that have interests similar to theirs.  We compute
weights given to each candidates based on these assumptions.

For example, suppose $w$ created a link to $v$, and there are multiple
candidate users $u_1,\ldots u_n$ as the user imitated by $w$.
Figure~\ref{fig:rival} illustrates this situation.  If the interests
of $u_i$ and $v$ are very similar, $v$ is likely to be an information
source on a topic for which $u_i$ is a good early adopter.  Similarly,
if the interests of $w$ and $u_i$ are similar, $w$ is more likely to
imitate $u_i$ than other $u_j$ whose interests are not similar to
interests of $w$.

We measure similarity between interests of two users by the similarity
of their sets of friends.  Similarity between $u, v\in V$, denoted by
$\mathit{Sim}(u, v)$, is defined as follows:
\begin{dfn}
Similarity between interests of $u$ and $v$:
\[
\mathit{Sim}(u, v) = \frac{|\mathit{Friends}(u) \cap \mathit{Friends}(v)|}{|\mathit{Friends}(u)
\cup \mathit{Friends}(v)|}
\label{eq:sim}
\]
\end{dfn}

The details of how to assign weighted probability to candidates is
explained in Section~\ref{EstimateCopied}.

\subsection{Estimation of Imitation Frequency}
\label{EstimateCopied}

Now we explain how we estimate $|\mathit{Copy}(u)|$, imitation
frequency of a user $u$, with including network structure and the three
optional factors above.  We first estimate the probability that a
follow link $\langle w, v\rangle$ was created by imitating the link
$\langle u, v\rangle$, denoted by $P_{\langle w,v\rangle}(u)$, by the
formula below:
\begin{dfn}
The probability that the link $\langle w, v\rangle\in E$ is a copy of
$\langle u, v\rangle\in E$:
\[
P_{\langle w, v\rangle}(u) = \frac{p_{\langle w,v\rangle}(u)}{\sum_{x \in V}p_{\langle w,v\rangle}(x)}
\]
where
\[
\begin{split}
p_{\langle w, v \rangle}(x) = &\mathit{Structure(x, v, w)} \times \mathit{Time(x, v, w)} \times\\
& \mathit{Nonrec}(x,v,w) \times \mathit{Sim}(x, v) \times \mathit{Sim}(w, x)
\end{split}
\]
\end{dfn}

The formula above corresponds to the case where we use all three
optional factors.  When we do not use some of them, we simply remove
terms corresponding to them from the formula.  The Boolean values of
$\mathit{Structure}(x, v, w)$, $\mathit{Time}(x, v, w)$, and
$\mathit{Nonrec}(v, u, w)$ are interpreted as 1 or 0, and they are
used to give the score 0 to accounts that cannot be the one that were
imitated.  $\mathit{Sim}(x, v)$ and $\mathit{Sim}(w, x)$ are used to
give proper weights to multiple candidates.

We then estimate $|\mathit{Copy}(u)|$, required for the computation of
the final early adopter scores,
as follows:
\begin{dfn}
$\mathit{CF}(u)$, the expected value of $|\mathit{Copy}(u)|$:
\[
\mathit{CF}(u) = \sum_{\langle w,v\rangle\in E}P_{\langle w,v\rangle}(u)
\]
\end{dfn}
We estimate the expected value of the number of times $u$ was imitated
by summing the probability that each candidate link is a copy of the
link of $u$.

\subsection{Algorithms}

We have designed two algorithms to compute early adopter scores of the
followers of the given new accounts.  The first one computes early
adopter scores of all accounts in the graph by exactly following the
definition above.  For each follow link in the given graph, we collect
candidate links that can be the original of the link, and give the
owner of each link the probability that it is the original.  At each
account, these given probabilities are accumulated.  By summing up all
these probability values given to a user $u$, we obtain the expected
value of $|\mathit{Copy}(u)|$.  This algorithm is shown in
Algorithm~\ref{algo:global}.

This algorithm evaluates $c_{\langle w,v\rangle}(u)$ for $O(md)$ times
where $m$ is the number of edges in the graph and $d$ is the degree of
nodes.  When we use no optional factors, we actually do not need to
compute $c_{\langle w,v\rangle}(u)$ in Algorithm~\ref{algo:global}
because it always returns 1 given that $\langle w,v\rangle\in E$ and
$u\in\mathit{Followers}(v)\cap\mathit{Friends}(w)$.  Therefore, the
time complexity of this algorithm is in $O(md)$ in that case, if we
assume $\mathit{Followers}(u)$ and/or $\mathit{Friends}(u)$ are stored
in a hash table.  Even if we include the factor $\mathit{Nonrec}$, we
can simply skip the loop for such $\langle w,v\rangle$, and the
complexity of the algorithm is still in $O(md)$.  According to our
experiments, which will be explained later, the other two factors are
not actually useful, so the time complexity of our best method is in
$O(md)$.

\begin{algorithm}[t]
\caption{Computing Early Adopter Scores of All Nodes}
\label{algo:global}
\begin{algorithmic}
\STATE $\mathit{CF}(u):=0$ for all $u\in V$
\FORALL{$\langle w,v\rangle\in E$}
\STATE $U:=\mathit{Followers}(v)\cap\mathit{Friends}(w)$
\STATE $s:=0$
\FORALL{$u\in U$}
\STATE $c[u]:=c_{\langle w,v\rangle}(u)$
\STATE $s:=s+c[u]$
\ENDFOR
\FORALL{$u\in U$}
\STATE $CF(u) = CF(u) + c[u]/s$
\ENDFOR
\ENDFOR
\RETURN $CF(u)$ for all $u\in V$
\end{algorithmic}
\end{algorithm}

The algorithm above computes early adopter scores for all nodes, but
when we only want to compute a future popularity score of one new
account, we only need to compute early adopter scores of its
followers.  For such cases, we designed another algorithm to compute
the early adopter score of a given user $u$.  We omit the details, but
it simply collects all candidate triadic closures by retrieving all
the friends of the followers of $u$, and checking if they are friends
of $u$.  Its time complexity is in $O(d^3)$, and therefore we can
compute the future popularity score of a given account in $O(d^4)$.
According to our experiment, however, this algorithm can be slower
even when we need to compute early adopter scores only for less than a
thousand of nodes.

\subsection{Extension to a Recursive Method}

The method explained above computes the future popularity score of an
account based on the early adopter scores of its direct followers.  We
can easily extend this method to a recursive method based on various
infection models.

As explained before, $E_1(u,v)$ represents the expected number of link
propagation from $u$ to its followers and $E_2(u,v)$ represents the
probability that links are propagated from $u$ to its followers.
We can interpret them as the propagation probability of a disease, and
can run some algorithms that predict how many users will be infected
starting from a given infected user.  We tested such recursive versions
of our method by using some simple algorithms, in our experiment, such
recursive method did not improve the performance of our method.  We
will investigate this problem in our future research.

\section{Experiment}
\label{sec:experiment}

In this section, we evaluate our method by the experiment on the
Twitter data set.  We first explain the data set used in our
experiment and the procedure of our experiment.  After that, we will
explain the baseline methods with which we compared our methods.
Finally, we show and discuss the results of the experiment.

\subsection{Data Set}

We use the snapshot of a part of Twitter follow graph created by Rui
et al.\ in May 2011 \cite{DBLP:conf/kdd/LiWDWC12}.  This data set was
produced by random crawling of follow links starting from randomly
selected 100,000 users.  In this graph, $|V| = 21,604,165$ and
$|E| = 284,885,001$.
We denote this network by $D_{11}(V, E)$ in order to distinguish it
from another network explained later.

We extracted all accounts in $D_{11}$ that were within two weeks,
three weeks, and four weeks from its creation date, and that had at
least 10 followers, 20 followers, and 30 followers at the time of
$D_{11}$.  Let $T^2_{10}$, $T^2_{20}$, $T^2_{30}$, $T^3_{10}$,
$T^3_{20}$, $T^3_{30}$, $T^4_{10}$, $T^4_{20}$, $T^4_{30}$ denote
these data sets.  Therefore, $T^{x}_{30} \subseteq T^{x}_{20}
\subseteq T^{x}_{10}$ and $T^{2}_{x} \subseteq T^{3}_{x} \subseteq
T^{4}_{x}$.  Their size is shown at the top of
Table~\ref{table:correlation}.

\subsection{Procedure of Experiment}
\label{ExDetail}

We run our experiment in the following procedure:

\begin{enumerate}
\item For all accounts in the data set, we estimated their future
popularity by our methods and by various baseline methods, and produce
a list of accounts sorted in the order of their estimated future
popularity.
\item We used the number of their non-reciprocal followers as of May
2015, which we denote $\mbox{FW}_{\mathit{nr}}^{2015}(u)$, as the true future
popularity of the information sources, and produce a list of accounts
sorted in that order.
\item We compare the list produced by each estimation method and the
list based on $\mbox{FW}_{\mathit{nr}}^{2015}(u)$.  For the comparison, we
used Spearman's rank correlation coefficient ($\rho$) and the
normalized discount cumulative gain (nDCG).  Spearman's $\rho$
reflects the accuracy of the whole estimated ranking, while nDCG only
reflects the accuracy of the top part of the estimated ranking.
\end{enumerate}

\begin{table*}[t]
\begin{center}
\small
\begin{tabular}{lll|rrr|rrr|rrr|rrr|rrr}
\hline
\multicolumn{3}{r|}{data set} &$T^{4}_{10}$&$T^{3}_{10}$&$T^{2}_{10}$&$T^{4}_{20}$&$T^{3}_{20}$&$T^{2}_{20}$&$T^{4}_{30}$&$T^{3}_{30}$&$T^{2}_{30}$&$\hat{T}^{4}_{10}$&$\hat{T}^{3}_{10}$&$\hat{T}^{2}_{10}$&$\hat{T}^{4}_{20}$&$\hat{T}^{3}_{20}$&$\hat{T}^{2}_{20}$\\
\multicolumn{3}{r|}{data size\rule{0mm}{8pt}} & 6921 & 3270 & 1515 & 2259 & 1005 & 431 & 979 & 396 & 165 & 2249 & 1009 & 415 & 709 & 314R & 123\\
\hline
&\multicolumn{2}{l|}{$\mbox{FW}$}        & 0.18 & 0.20 & 0.23 & 0.15 & 0.12 & 0.07 & 0.19 & 0.18 & 0.00 & 0.19 & 0.19 & 0.19 & 0.20 & 0.20 & 0.19 \\
&\multicolumn{2}{l|}{$\mbox{FW}_{\mathit{nr}}$}    & 0.11 & 0.07 & 0.08 & 0.02 & 0.01 &-0.01 & 0.00 &-0.03 &-0.01 & 0.07 & 0.10 & 0.08 & 0.00 & 0.03 &-0.03 \\
&\multicolumn{2}{l|}{$\mbox{FR}$}        & 0.19 & 0.25 & {\bf 0.31} & 0.24 & 0.20 & 0.22 & 0.33 & 0.30 & 0.35 & 0.26 & 0.24 & 0.23 & 0.34 & 0.37 & 0.53 \\
&\multicolumn{2}{l|}{$\mbox{FR}_{\mathit{nr}}$}    & 0.04 &-0.05 &-0.04 &-0.05 & 0.01 & 0.04 &-0.05 &-0.04 &-0.05 & 0.02 & 0.09 & 0.13 &-0.09 & 0.02 & 0.08 \\
&\multicolumn{2}{l|}{$\mbox{HITS}$}      & 0.15 & 0.11 & 0.10 & 0.11 & 0.13 & 0.10 & 0.04 & 0.07 & 0.08 & 0.13 & 0.16 & 0.12 & 0.12 & 0.11 & 0.02 \\
&\multicolumn{2}{l|}{$\mbox{HITS}_{\mathit{nr}}$}  & {\bf 0.26} & {\bf 0.27} & {\bf 0.31} & {\bf 0.30} & 0.30 & 0.35 & {\bf 0.38} & \underline{\bf 0.38} & 0.46 & {\bf 0.33} & {\bf 0.31} & {\bf 0.33} & {\bf 0.41} & 0.45 & {\bf 0.61}\\
&\multicolumn{2}{l|}{$\mbox{PR}$}        & 0.20 & 0.15 & 0.13 & 0.20 & 0.15 & 0.14 & 0.24 & 0.19 & 0.25 & 0.16 & 0.14 & 0.14 & 0.21 & 0.19 & 0.26 \\
&\multicolumn{2}{l|}{$\mbox{PR}_{\mathit{nr}}$}    & 0.16 & 0.12 & 0.09 & 0.21 & 0.17 & 0.20 & 0.30 & 0.27 & 0.32 & 0.21 & 0.19 & 0.22 & 0.30 & 0.35 & 0.51 \\
&\multicolumn{2}{l|}{$\mbox{AD}_{\Sigma}$} &-0.06 &-0.08 &-0.08 &-0.17 &-0.21 &-0.27 &-0.18 &-0.24 &-0.26 &-0.12 &-0.10 &-0.15 &-0.22 &-0.28 &-0.40 \\
&\multicolumn{2}{l|}{$\mbox{AD}_{\mu}$}   &-0.21 &-0.17 &-0.13 & {\bf -0.30} & {\bf -0.38} & \underline{\bf -0.46} &-0.27 &-0.37 & \underline{\bf -0.50} &-0.27 & {\bf -0.31} &-0.31 &-0.35 & {\bf -0.50} &-0.54 \\
\hline
\rule{0cm}{1em}
$F_1$ & $\Sigma$ & -   & 0.30 & 0.27 & 0.29 & 0.24 & 0.25 & 0.27 & 0.23 & 0.23 & 0.32 & 0.28 & 0.32 & 0.36 & 0.29 & 0.32 & 0.40 \\
      &          & r   & 0.36 & 0.35 & 0.38 & 0.32 & 0.32 & 0.38 & 0.34 & 0.36 & 0.46 & 0.35 & 0.36 & 0.39 & 0.38 & 0.42 & 0.57 \\
      &          & s   & 0.30 & 0.26 & 0.28 & 0.24 & 0.25 & 0.27 & 0.21 & 0.22 & 0.31 & 0.28 & 0.32 & 0.36 & 0.28 & 0.31 & 0.42 \\
      &          & r s & 0.35 & 0.33 & 0.36 & 0.29 & 0.30 & 0.36 & 0.32 & 0.33 & 0.44 & 0.32 & 0.33 & 0.38 & 0.34 & 0.38 & 0.55 \\
\cline{2-18}
\cline{2-18}
      & g        & -   & 0.30 & 0.28 & 0.31 & 0.27 & 0.27 & 0.31 & 0.25 & 0.24 & 0.35 & 0.29 & 0.31 & 0.36 & 0.31 & 0.34 & 0.43 \\
      &          & r   & 0.35 & 0.34 & 0.37 & 0.31 & 0.32 & 0.36 & 0.33 & 0.35 & 0.46 & 0.35 & 0.35 & 0.38 & 0.36 & 0.41 & 0.53 \\
      &          & s   & 0.30 & 0.28 & 0.30 & 0.27 & 0.27 & 0.30 & 0.24 & 0.22 & 0.29 & 0.29 & 0.31 & 0.37 & 0.30 & 0.33 & 0.44 \\
      &          & r s & 0.35 & 0.32 & 0.35 & 0.31 & 0.32 & 0.37 & 0.32 & 0.33 & 0.43 & 0.33 & 0.33 & 0.37 & 0.35 & 0.37 & 0.52 \\
\cline{1-18}
\rule{0cm}{1em}
$F_2$ & $\Sigma$ & -   & \underline{\bf 0.39} & 0.37 & 0.39 & 0.36 & 0.37 & 0.42 & 0.34 & 0.32 & 0.40 & 0.39 & 0.40 & 0.44 & 0.44 & 0.50 & 0.62 \\
      &          & r   & \underline{\bf 0.39} & \underline{\bf 0.39} & \underline{\bf 0.41} & \underline{\bf 0.39} & \underline{\bf 0.39} & {\bf 0.45} & \underline{\bf 0.40} & \underline{\bf 0.38} & {\bf 0.47} & 0.39 & 0.39 & 0.43 & \underline{\bf 0.46} & \underline{\bf 0.52} & \underline{\bf 0.64}\\
      &          & s   & 0.38 & 0.37 & 0.38 & 0.36 & 0.36 & 0.41 & 0.33 & 0.31 & 0.38 & 0.38 & 0.39 & 0.43 & 0.43 & 0.48 & 0.61\\
      &          & r s & \underline{\bf 0.39} & 0.38 & 0.40 & 0.38 & 0.38 & {\bf 0.45} & 0.39 & 0.37 & {\bf 0.47} & 0.38 & 0.38 & 0.42 & 0.45 & 0.50 & 0.64 \\\cline{2-18}
      & g        & -   & 0.38 & 0.36 & 0.37 & 0.35 & 0.36 & 0.39 & 0.31 & 0.30 & 0.37 & \underline{\bf 0.42} & \underline{\bf 0.45} & \underline{\bf 0.47} & 0.43 & 0.49 & 0.58 \\
      &          & r   & \underline{\bf 0.39} & 0.38 & \underline{\bf 0.41} & 0.38 & 0.38 & 0.44 & 0.38 & 0.37 & 0.45 & 0.39 & 0.39 & 0.43 & 0.45 & 0.51 & 0.63 \\
      &          & s   & 0.38 & 0.36 & 0.36 & 0.35 & 0.36 & 0.40 & 0.31 & 0.31 & 0.38 & \underline{\bf 0.42} & \underline{\bf 0.45} & \underline{\bf 0.47} & 0.42 & 0.46 & 0.58 \\
      &          & r s & \underline{\bf 0.39} & 0.38 & 0.40 & 0.38 & 0.38 & 0.44 & 0.38 & 0.36 & 0.45 & 0.38 & 0.39 & 0.42 & 0.45 & 0.50 & 0.63 \\
\hline
\multicolumn{3}{r|}{\rule{0cm}{1em}$\mbox{LR}$} & 0.43 & 0.43 & 0.46 & 0.43 & 0.45 & 0.50 & 0.45 & 0.46 & 0.58 & 0.47 & 0.50 & 0.50 & 0.52 & 0.57 & 0.65 \\
\hline
\end{tabular}
\caption{Spearman's $\rho$ between $\mbox{FW}^{2015}_{\mathit{nr}}$
and each method.  $\mbox{FW}$ to $\mbox{AD}_{\mu}$ are baseline
methods, $F_1$ and $F_2$ are our methods, and $LR$ is the logistic
regression combining some baseline methods and our methods.  $T$
denotes data sets including both active and non-active users.
$\hat{T}$ denotes data sets only including active users.  For each
data set, the best scores among the baseline methods and the best
score among 16 variations of our method are shown in bold fonts.  The
best scores among both of them are also underlined.  A variation of
our method $F_2^{\Sigma}(r)$ outperforms the baseline methods except
for $T^2_{20}$ and $T^2_{30}$.  Notice that $F_2^{\Sigma}(r)$
outperforms the baseline methods even for $\hat{T}^{x}_{10}$, where
the score of $F_2^{\Sigma}(r)$ are not in bold fonts simply because
$F_2^{g}$ was the best among our methods.}
\label{table:correlation}
\end{center}
\end{table*}

\subsection{Tested Proposed Methods}

In Section~\ref{sec:propagation}, we showed two definitions of early
adopter scores, $E_1(u,v)$ and $E_2(u,v)$, and we also showed two ways
to calculate future popularity scores, $F_i^{\Sigma}$ and $F_i^g$.  We
also have three optional factors in the estimation of $|\mathit{Copy}(u)|$, and
there are eight combinations of them.  In total, we have 32
combinations of them and we compared them in our experiment.  In this
paper, however, we omit the result of the methods that use temporal
order of links because it did not improve the accuracy of our method,
and also because of the spece limitation.

In the following, $r$ denotes the option of link reciprocity, and $s$
denote the option of similarity between users.  For example,
$F_2^{\Sigma}(r)$ represents our method using $E_2(u,v)$,
$F_i^{\Sigma}(v)$, and only link reciprocity option.

The parameter $c$ for g-index was hand-tuned to the following values in
each case.  $E_1$: 50000, $E_1+r$: 100000, $E_1+s$: 50000, $E_1+r,s$:
50000, $E_2$: 1, $E_2+r$: 10, $E_2+s$: 1, $E_2+r,s$: 10.

\subsection{Baseline Methods}

We next explain the baseline methods we compared and their parameters.

\noindent
\textbf{Followers ($\mbox{FW}$)}: It measures the future popularity of
new accounts by the number of their current followers in May 2011.

\noindent
\textbf{Nonreciprocal followers ($\mbox{FW}_{\mathit{nr}}$)}: It measures it by
the number of their non-reciprocal followers in May 2011.  As
explained in Section~\ref{subsec:nonreciprocalFollow}, non-reciprocal
follow links are likely to be links to information sources.

\noindent
\textbf{Friends ($\mbox{FR}$)}: It measures it by the number of friends
in May 2011.

\noindent
\textbf{Nonreciprocal friends ($\mbox{FR}_{\mathit{nr}}$)}: It measures it by
the number of their non-reciprocal friends in May 2011.

\noindent
\textbf{HITS}: It computes authority scores and hub scores of accounts
\cite{romero2010directed}, and use the authority score as the
indicator of future popularity.  In this experiment, we set the number
of iterations to 10, with which the scores sufficiently converged.

\noindent
\textbf{Nonreciprocal HITS ($\mbox{HITS}_{\mathit{nr}}$)}: The same as HITS,
but it computes authority and hub score on the graph consisting only
of non-reciprocal links.  The number of iterations is 10, with which
the scores sufficiently converged.

\noindent
\textbf{PageRank ($\mbox{PR}$)}: It estimates the future popularity by
using PageRank score \cite{BrynPageRank}. In this experiment, we set
the damping factor $d = 0.9$ and number of iterations to 100, with
which the scores sufficiently converged.

\noindent
\textbf{Nonreciprocal PageRank ($\mbox{PR}_{\mathit{nr}}$)}: The same
as PageRank, but it computes PageRank scores on the graph consisting
only of nonreciprocal links.  We set the damping factor $d = 0.9$ and
the number of iterations is 100, with which the scores sufficiently
converged.

\noindent
\textbf{Adamic/Adar ($\mbox{AD}_\Sigma$, $\mbox{AD}_\mu$)}: It
estimates the future popularity of $v$ by estimating the probability
of new links to $v$ from other nodes based on Adamic/Adar index
\cite{Adamic01friendsand}.  Given an account $v$, we collect all its
friends, and also all the followers of those friends.  Then we compute
Adamic/Adar index for $v$ and all these followers with regarding their
common friends as the common items.  The ordinary Adamic/Adar sums all
the obtained index values, but we compared both summation and average.

\subsection{Result and Discussion}

We next analyze and discuss the results of our experiment.

We first analyze the correlation between the ranking lists based on
$\mbox{FW}_{\mathit{nr}}^{2015}$ and each method by Spearman's $\rho$.
The left half of Table~\ref{table:correlation} lists the $\rho$ values
between $\mbox{FW}_{\mathit{nr}}^{2015}$ and each method.  In each
column in Table~\ref{table:correlation}, the best scores among the
baseline methods and the best score among the 16 variations of our
method are shown in bold fonts.

Among the baselines, $\mbox{FW}$ and $\mbox{FR}$ achieved higher
correlation than those without reciprocal links.  On the contrary,
$\mbox{HITS}$ and $\mbox{PR}$ without reciprocal links achieved higher
correlation than those with reciprocal links.  Among them,
$\mbox{HITS}_{\mathit{nr}}$ was the best, and $\mbox{PR}_{\mathit{nr}}$ follows.
These methods have higher correlation for the data set including
accounts with more followers.

On the other hand, AD methods have negative correlation, and
AD$_{\mu}$ has, surprisingly, high negative correlation, which means
it is a good index for predicting future popularity of accounts.  It
also has higher correlation for the data set including accounts with
more followers.

However, our method, especially $F_{2}^{\Sigma}(r)$ achieves even
higher correlation except for two cases for the data set $T^2_{20}$
and $T^2_{30}$.  Our method is outperformed by AD$_{\mu}$ in these
cases, where the data set includes accounts that have already obtained
many followers (more than 20 or 30) within a short time (2 weeks).
This suggests that our method is especially good at the discovery of
accounts that start with a fewer followers, but then become popular
later.

We performed some error-analysis, and the result shows that the main
factor lowering the accuracy of all the compared methods is the
existence of many accounts that had some followers in $D_{11}$ but are
inactive or deleted as of 2015.  Therefore, we also created data sets
$\hat{T}$ that include only accounts that are active as of 2015.  The
right half of Table~\ref{table:correlation} shows the result on these
data sets.  In thes data sets, we again outperforms baselines, and
achieves higher accuracy than for $T$, i.e., the data set including
inactive users.  This suggest that we should combine our method with
some method that predict if a given account will last long or not.
Notice that $F_2^{\Sigma}(r)$ outperforms the baseline methods in all
cases although their scores are not in bold fonts in some cases where
they are outperformed by $F_2^{g}$.

\begin{figure}[t]
  \centering
  \includegraphics[width=4.1cm,bb=0 0 489 462]{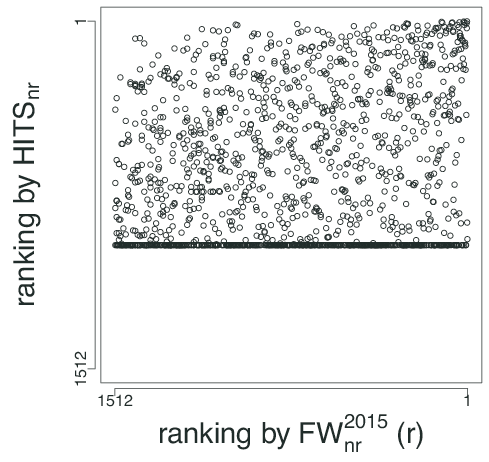}
  \includegraphics[width=4.1cm,bb=0 0 489 462]{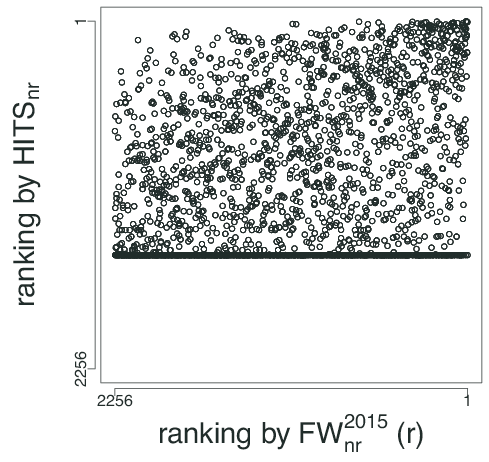}
  \includegraphics[width=4.1cm,bb=0 0 489 462]{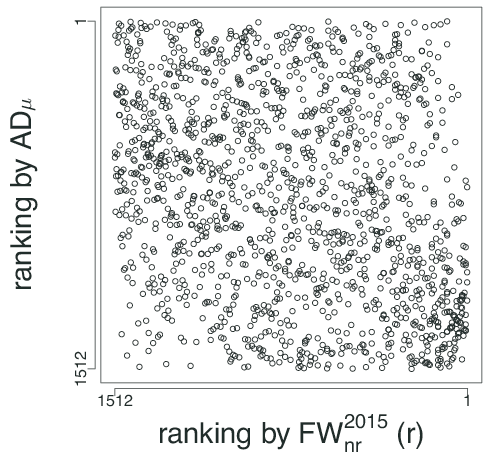}
  \includegraphics[width=4.1cm,bb=0 0 489 462]{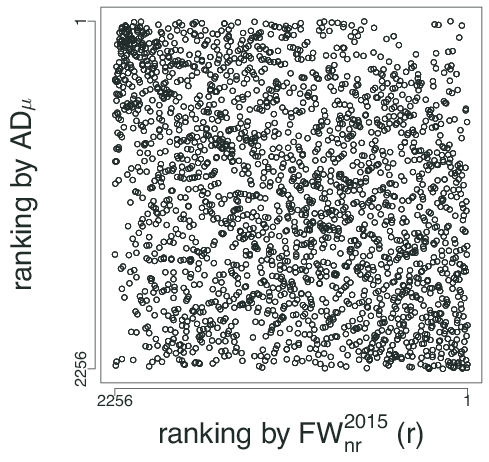}
  \includegraphics[width=4.1cm,bb=0 0 489 462]{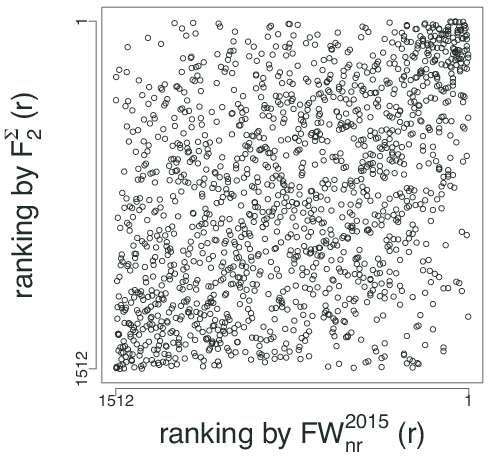}
  \includegraphics[width=4.1cm,bb=0 0 489 462]{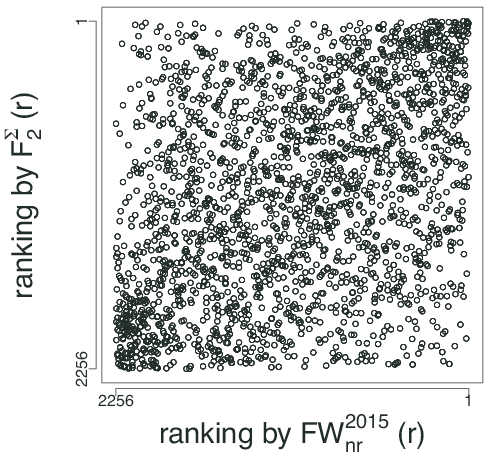}
 \caption{Scatter diagram for $\mbox{HITS}_{\mathit{nr}}$,
 $\mbox{AD}_\mu$, $F_2^{\Sigma}(r)$, which had high $\rho$
 value, for $T^{2}_{10}$ (left) and $T^{4}_{20}$ (right).
The horizontal row of points near the bottom in the daigram for
$\mbox{HITS}_{\mathit{nr}}$ corresponds to the accounts that had no
non-reciprocal followers in $D_{11}$.}
 \label{plotOfRankingCorrelative20}
\end{figure}

Figure~\ref{plotOfRankingCorrelative20} show scatter diagrams between
$\mbox{FW}_{\mathit{nr}}^{2015}$ and each method that showed high correlation,
i.e., $\mbox{HITS}_{\mathit{nr}}$, $\mbox{AD}_\mu$, $F_2^{\Sigma}(r)$,
for the data set $T^{2}_{10}$ (left) and $T^{4}_{20}$ (right).  In the
scatter diagrams for
$\mbox{HITS}_{\mathit{nr}}$,
there are a horizontal row of points near the bottom. They are
accounts that had no non-reciprocal followers in $D_{11}$.  The
methods that use non-reciprocal links achieves higher $\rho$ values,
but they also have this problem.

\begin{table*}[t]
 \begin{center}
\small
\begin{tabular}{lll|rrrr|rrrr|rrrr}
\hline
\multicolumn{3}{c|}{data set} & \multicolumn{4}{c|}{$T^4_{10}$} & \multicolumn{4}{c|}{$T^3_{10}$} & \multicolumn{4}{c}{$T^2_{10}$} \\
\multicolumn{3}{c|}{nDCG@k}              & $@10$ & $@20$ & $@50$ & $@100$ & $@10$ & $@20$ & $@50$ & $@100$ & $@10$ & $@20$ & $@50$ & $@100$ \\
\hline
\rule{0cm}{1em}
&\multicolumn{2}{l|}{$\mbox{FW}$}        & 0.05 & 0.04 & 0.05 & 0.05 & 0.19 & 0.17 & 0.15 & 0.18 & 0.03 & 0.06 & 0.11 & 0.11 \\
&\multicolumn{2}{l|}{$\mbox{FW}_{\mathit{nr}}$}    & 0.08 & 0.15 & 0.14 & 0.16 & 0.07 & 0.07 & 0.14 & 0.15 & 0.17 & 0.17 & 0.18 & 0.18 \\
&\multicolumn{2}{l|}{$\mbox{FR}$}        & 0.02 & 0.02 & 0.03 & 0.03 & 0.17 & 0.15 & 0.14 & 0.14 & 0.01 & 0.01 & 0.03 & 0.05 \\
&\multicolumn{2}{l|}{$\mbox{FR}_{\mathit{nr}}$}    & 0.00 & 0.00 & 0.01 & 0.01 & 0.00 & 0.00 & 0.00 & 0.01 & 0.00 & 0.00 & 0.00 & 0.01 \\
&\multicolumn{2}{l|}{$\mbox{HITS}$}      & 0.00 & 0.02 & 0.02 & 0.03 & 0.11 & 0.10 & 0.09 & 0.10 & 0.01 & 0.01 & 0.02 & 0.03 \\
&\multicolumn{2}{l|}{$\mbox{HITS}_{\mathit{nr}}$}  & \underline{\bf 0.19} & \underline{\bf 0.19} & \underline{\bf 0.18} & \underline{\bf 0.19} & 0.13 & 0.14 & 0.18 & 0.20 & \underline{\bf 0.25} & \underline{\bf 0.27} & \underline{\bf 0.29} & \underline{\bf 0.30} \\
&\multicolumn{2}{l|}{$\mbox{PR}$}        & 0.07 & 0.07 & 0.07 & 0.13 & \underline{\bf 0.23} & \underline{\bf 0.24} & {\bf 0.23} & {\bf 0.24} & 0.12 & 0.12 & 0.17 & 0.20 \\
&\multicolumn{2}{l|}{$\mbox{PR}_{\mathit{nr}}$}    & 0.17 & 0.16 & \underline{\bf 0.18} & 0.18 & 0.02 & 0.06 & 0.11 & 0.15 & 0.11 & 0.14 & 0.16 & 0.17 \\
&\multicolumn{2}{l|}{$\mbox{AD}_{\Sigma}$} & 0.07 & 0.13 & 0.14 & 0.14 & 0.20 & 0.22 & 0.21 & 0.20 & \underline{\bf 0.25} & 0.24 & 0.23 & 0.24 \\
&\multicolumn{2}{l|}{$\mbox{AD}_{\mu}$}   & 0.00 & 0.00 & 0.00 & 0.01 & 0.00 & 0.00 & 0.01 & 0.02 & 0.00 & 0.01 & 0.02 & 0.06 \\
\hline
\rule{0cm}{1em}
$F_2$ & $\Sigma$ & -   & 0.03 & 0.05 & 0.09 & 0.11 & {\bf 0.21} & {\bf 0.18} & \underline{\bf 0.25} & \underline{\bf 0.25} & 0.13 & 0.16 & 0.16 & 0.16 \\
      &          & r   & 0.00 & 0.00 & 0.00 & 0.00 & 0.00 & 0.00 & 0.00 & 0.00 & 0.00 & 0.00 & 0.01 & 0.06 \\
      &          & s   & {\bf 0.06} & 0.06 & 0.05 & 0.05 & 0.19 & 0.16 & 0.15 & 0.16 & 0.03 & 0.03 & 0.04 & 0.09 \\
      &          & r,s & {\bf 0.06} & {\bf 0.13} & {\bf 0.14} & {\bf 0.15} & 0.12 & 0.14 & 0.17 & 0.17 & {\bf 0.22} & {\bf 0.21} & {\bf 0.20} & {\bf 0.22} \\
\cline{2-15}
\cline{2-15}
      & $g$      & -   & 0.02 & 0.02 & 0.02 & 0.02 & 0.02 & 0.01 & 0.02 & 0.05 & 0.03 & 0.03 & 0.04 & 0.11 \\
      &          & r   & {\bf 0.06} & 0.11 & 0.13 & 0.14 & 0.07 & 0.10 & 0.13 & 0.15 & 0.16 & 0.19 & 0.18 & 0.19 \\
      &          & s   & 0.00 & 0.00 & 0.01 & 0.02 & 0.01 & 0.01 & 0.03 & 0.03 & 0.02 & 0.05 & 0.05 & 0.08 \\
      &          & r,s & 0.05 & 0.04 & 0.04 & 0.04 & 0.03 & 0.03 & 0.03 & 0.03 & 0.03 & 0.03 & 0.03 & 0.03 \\
\hline
\end{tabular}
\caption{ $\mbox{nDCG}@k$ by each methods where gain is
$\mbox{FW}_{\mathit{nr}}^{2014}$.  Each column shows values of $k$, and the
bold entries are highest score among the baseline methods and among
the variations of our method.  The best scores among both of them are
also underlined.}
\label{table:ndcg}
\end{center}
\end{table*}

\begin{table*}[t]
 \begin{center}
\small
  \begin{tabular}{l|r|r|r|r|r|r|r|r|r|r|r|r|r|r}
&$\mbox{FW}$&$\mbox{FW}_{\mathit{nr}}$&$\mbox{FR}$&$\mbox{FR}_{\mathit{nr}}$&$\mbox{HITS}$&$\mbox{HITS}_{\mathit{nr}}$&$\mbox{PR}$&$\mbox{PR}_{\mathit{nr}}$&$\mbox{AD}_{\Sigma}$&$\mbox{AD}_{\mu}$&$F_2^{\Sigma}$&$F_2^{\Sigma}(r)$&$F_2^{\Sigma}(s)$&$F_2^{\Sigma}(r,s)$\\ \hline \rule{0cm}{1em}
$\mbox{FW}$          & 1.00 &  0.16&   0.69 &  0.18 &  0.57 &  0.13 &  0.32 &  0.19 & 0.65 & 0.31  & 0.52 & -0.01 & 0.60 & 0.35  \\
$\mbox{FW}_{\mathit{nr}}$     &     &  1.00&   -0.08&   0.03&   0.15&   0.82&   0.12&  0.87 &-0.03 & -0.08 & 0.08 & -0.02 & 0.09 & 0.10 \\
$\mbox{FR}$          &     &     &   1.00 &  0.71 &  0.57 &  -0.05&   0.36&  -0.03 & 0.78 & 0.27  & 0.47 & 0.11 & 0.51 & 0.36 \\
$\mbox{FR}_{\mathit{nr}}$     &     &     &       &  1.00 &  0.32 &  0.05 &  0.25 &  0.04 & 0.47 & 0.06  & 0.21 & 0.14 & 0.20 & 0.22  \\
$\mbox{HITS}$        &     &     &       &      &  1.00 &  0.07 &  0.35 &  0.16 & 0.64 &-0.07 & 0.22 & -0.11 & 0.30 & 0.05 \\
$\mbox{HITS}_{\mathit{nr}}$   &     &     &       &      &      &  1.00 &  0.15 &  0.74 &-0.15 & -0.18 & 0.28 & 0.25 & 0.26 & 0.34 \\
$\mbox{PR}$          &     &     &       &      &      &      &  1.00 &  0.12 & 0.20 &-0.06 & 0.37 & 0.21 & 0.37 & 0.37 \\
$\mbox{PR}_{\mathit{nr}}$     &     &     &       &      &      &      &      &  1.00 &-0.04 & -0.10 & 0.18 & 0.09 & 0.18 & 0.18  \\
$\mbox{AD}_{\Sigma}$    &     &     &       &      &      &      &      &       & 1.00 & 0.46 & 0.02 & -0.41 & 0.11 &-0.12 \\
$\mbox{AD}_{\mu}$   &     &     &       &      &      &      &      &       &     & 1.00 & -0.08 & -0.34 & -0.02 & -0.16 \\
$F_2^{\Sigma}$         &     &     &       &      &      &      &      &       &     &      & 1.00 &  0.81 & 0.97 & 0.93 \\
$F_2^{\Sigma}(r)$       &     &     &       &      &      &      &      &       &     &      &      & 1.00 & 0.71 & 0.85 \\
$F_2^{\Sigma}(s)$       &     &     &       &      &      &      &      &       &     &      &      &      & 1.00 & 0.87 \\
$F_2^{\Sigma}(r,s)$     &     &     &       &      &      &      &      &       &     &      &      &      &      &1.00 \\

  \end{tabular}
  \caption{ Spearman's $\rho$ between each methods.  Our best methods
  and the best baseline methods have low correlation, which suggests
  that they are complementary to each other.
}
 \label{table:corr-methods}
 \end{center}
\end{table*}

In Figure~\ref{plotOfRankingCorrelative20}, the diagram for
$\mbox{AD}_\mu$ and $T^4_{20}$ shows that they are mainly good at
distinguishing the least popular accounts while $F_2^{\Sigma}(r)$ is
good at both the most popular accounts and the least popular accounts.
In order to examine these aspects in more detail, we also compared
baseline methods and our methods by the normalized discount cumulative
gain (nDCG).  nDCG is a measure of ranking quality, and takes value in
the range of $[-1, 1]$. In nDCG, accuracy of the top part of a ranking
is more important than the lower part.
nDCG@$k$ is a measure that computes nDCG only for top $k$ in the
ranking.  We calculated nDCG@$k$ of each method with various $k$ for
some data set.
Table~\ref{table:ndcg} shows the result.  The best scores among the
baselines and those among our methods are shown in bold fonts, and the
best score among both of them are also underlined.

Among the baselines, $\mbox{HITS}_{\mathit{nr}}$, $\mbox{PR}_{\mathit{nr}}$, and
AD$_\mu$ have high scores for some cases.  In this comparison,
our method does not achieve as good performance as these baselines.
It is mainly because top-ranked accounts were already popular at 2,3,4
weeks after the creation.  This again suggests that our method is
mainly good at detecting accounts that are not popular now but will be
popular later.

We also calculated the correlation between our methods and baseline
methods.  Figure~\ref{table:corr-methods} shows the result.  This
result shows that there are very low correlation between the good
baseline methods, such as $\mbox{HITS}_{\mathit{nr}}$ and
$\mbox{AD}_{\mu}$, and our best methods.  This suggest that we can
achieve better performance by combining these methods.  Following this
observation, we tested the logistic regression combining the methods
that showed high correlation in this experiment.  We learned weight
parameter for each combined method and evaluated their results by
10-fold validation.  Table~\ref{table:linear-para} shows the result.
Both $F_{1}^{\Sigma}$ and $F_{2}^{\Sigma}$ were given high $\beta$
values, which means they highly contribute the result.
All $p$ values are small enough, which shows this result is
stasistically reliable.

The accuracy of this combined method is shown at the line LR at the
bottom of Table~\ref{table:correlation}.  This method achieves the
best accuracy in all cases.

\begin{table}[t]
 \begin{center}
\begin{tabular}{l|rr} \hline
Method\rule{0cm}{1em}         &\multicolumn{1}{c}{$\beta$}&\multicolumn{1}{c}{$p$}\\ \hline
$(Intercept)$                 & -1.23115                 & $< 2.0  \times 10^{-16}               $   \\
$\mbox{HITS}$                 & 0.46882                  & $1.21 \times 10^{-6}$   \\
$\mbox{HITS}_{\mathit{nr}}$            & 1.13858                  & $1.21 \times 10^{-11}$   \\
$\mbox{AD}_{\mu}$            & -1.01902                 & $< 2.00 \times 10^{-16}$   \\
$F_{1}^{\Sigma}r$      & 0.99312                 & $5.77 \times 10^{-15}$   \\
$F_{2}^{\Sigma}r$      & 0.98678                  & $2.40 \times 10^{-11}$   \\
\end{tabular}
\caption{Result of logistic regression \mbox{LR} for $T^{4}_{10}$.
Both $F_{1}^{\Sigma}r$ and $F_{2}^{\Sigma}r$ highly contribute to the result.}
\label{table:linear-para}
\end{center}
\end{table}

\section{Conclusion}
\label{sec:conclusion}

In this paper, we proposed a method of predicting future popularity of
new Twitter accounts.  Our approach is based on the concept of early
adopters.  Early adopters are users that can find new useful
information sources earlier than other users.  Even if a new account
currently has only a few followers, if the followers are good early
adopters, we expect the new account will have many followers in
future.  We find early adopters based on the frequency of link
imitation, i.e., how often their follow links are imitated by their
followers.  We, therefore, need information on who imitated which
links in order to find early adopters, but that information is not
immediately available from the current Twitter graph.  We developed a
method that infer it by using four factors: network structure,
temporal order of creation of links, similarity between interests of
users, and reciprocity of links.

We evaluated the performance of our approach by estimating future
popularity of new Twitter accounts, and comparing the result with the
number of followers they actually obtained later.  The results show
that our approach achieves higher accuracy in the prediction of future
popularity of new information sources than various baseline methods.
Our approach outperforms the baselines especially for users that were
not popular at the time of prediction.  It means our approach is more
useful when we want to find new information sources that are not
popular now but will be popular in future.  Because the result of our
method and the best baseline method have low correlation, we also
tested logistic regression combining our method and the best baseline
methods, and it achieves even higher accuracy.

In this paper, we focus on the identification of early adopters, but
another interesting issue is what kind of properties these early
adopters have, and what makes them good early adopters.  In future
work, we will analyze early adopters identified by our method, and
clarify why they can find new good information sources earlier than
others, and why they are imitated by many users.

\bibliographystyle{abbrv}
\bibliography{www16imamori}

\begin{thebibliography}{10}

\bibitem{Adamic01friendsand}
L.~Adamic and E.~Adar.
\newblock Friends and neighbors on the web.
\newblock {\em Social Networks}, 25:211--230, 2001.

\bibitem{BakshySocialInfluence}
E.~Bakshy, B.~Karrer, and L.~A. Adamic.
\newblock Social influence and the diffusion of user-created content.
\newblock In {\em Proc.~of ACM EC}, pages 325--334, 2009.

\bibitem{rss2.0}
T.~R.~A. Board.
\newblock Rss 2.0 specification.
\newblock \url{http://www.rssboard.org/rss-specification}.

\bibitem{egghe2006theory}
L.~Egghe.
\newblock Theory and practice of the g-index.
\newblock {\em Scientometrics}, 69(1):131--152, 2006.

\bibitem{GoyalDiscoveringLeaders}
A.~Goyal, F.~Bonchi, and L.~V. Lakshmanan.
\newblock Discovering leaders from community actions.
\newblock In {\em Proc.~of CIKM}, pages 499--508, 2008.

\bibitem{hopcroft2011will}
J.~Hopcroft, T.~Lou, and J.~Tang.
\newblock Who will follow you back?: reciprocal relationship prediction.
\newblock In {\em Proc.~of CIKM}, pages 1137--1146. ACM, 2011.

\bibitem{hu2012people}
H.~Hu and X.~Wang.
\newblock How people make friends in social networking sites---a microscopic
  perspective.
\newblock {\em Physica A: Statistical Mechanics and its Applications},
  391(4):1877--1886, 2012.

\bibitem{KleinbergHITS}
J.~M. Kleinberg.
\newblock Authoritative sources in a hyperlinked environment.
\newblock {\em Journal of the ACM}, 46(5):604--632, 1999.

\bibitem{kwak2010twitter}
H.~Kwak, C.~Lee, H.~Park, and S.~Moon.
\newblock What is twitter, a social network or a news media?
\newblock In {\em Proc.~of WWW Conf.}, pages 591--600. ACM, 2010.

\bibitem{DBLP:conf/kdd/LiWDWC12}
R.~Li, S.~Wang, H.~Deng, R.~Wang, and K.~C.-C. Chang.
\newblock Towards social user profiling: unified and discriminative influence
  model for inferring home locations.
\newblock In {\em Proc.~of KDD}, pages 1023--1031, 2012.

\bibitem{liben2007link}
D.~Liben-Nowell and J.~Kleinberg.
\newblock The link-prediction problem for social networks.
\newblock {\em JASIST}, 58(7):1019--1031, 2007.

\bibitem{nguyen2010you}
V.-A. Nguyen, E.-P. Lim, H.-H. Tan, J.~Jiang, and A.~Sun.
\newblock Do you trust to get trust? a study of trust reciprocity behaviors and
  reciprocal trust prediction.
\newblock In {\em Proc.~of SDM}, pages 72--83, 2010.

\bibitem{BrynPageRank}
L.~Page, S.~Brin, R.~Motwani, and T.~Winograd.
\newblock The {PageRank} citation ranking: Bringing order to the web.
\newblock Technical Report 1999-66, Stanford InfoLab, November 1999.
\newblock Previous number = SIDL-WP-1999-0120.

\bibitem{rapoport1953spread}
A.~Rapoport.
\newblock Spread of information through a population with socio-structural
  bias: I. assumption of transitivity.
\newblock {\em The bulletin of mathematical biophysics}, 15(4):523--533, 1953.

\bibitem{romero2010directed}
D.~M. Romero and J.~M. Kleinberg.
\newblock The directed closure process in hybrid social-information networks,
  with an analysis of link formation on twitter.
\newblock In {\em Proc.~of ICWSM}, 2010.

\bibitem{SaezTrumperTrendsetters}
D.~Saez-Trumper, G.~Comarela, V.~Almeida, R.~Baeza-Yates, and F.~Benevenuto.
\newblock Finding trendsetters in information networks.
\newblock In {\em Proc.~ of SIGKDD}, pages 1014--1022, 2012.

\bibitem{tol2008rational}
R.~S. Tol.
\newblock A rational, successive g-index applied to economics departments in
  ireland.
\newblock {\em Journal of Informetrics}, 2(2):149--155, 2008.

\bibitem{weng2010twitterrank}
J.~Weng, E.-P. Lim, J.~Jiang, and Q.~He.
\newblock Twitterrank: finding topic-sensitive influential twitterers.
\newblock In {\em Proc.~of WSDM}, pages 261--270, 2010.

\bibitem{XieFriendTwitterLife}
W.~Xie, C.~Li, F.~Zhu, E.-P. Lim, and X.~Gong.
\newblock When a friend in twitter is a friend in life.
\newblock In {\em Proc.~of WebSci}, pages 344--347, 2012.

\bibitem{zhang2013learning}
J.~Zhang, C.~Wang, P.~S. Yu, and J.~Wang.
\newblock Learning latent friendship propagation networks with interest
  awareness for link prediction.
\newblock In {\em Proc.~of SIGIR}, pages 63--72. ACM, 2013.

\end{thebibliography}
\end{document}